\documentclass[12pt,aps]{revtex4}
\usepackage{epsf}
\usepackage{amsmath,amsthm,amssymb}
\usepackage{color}
\usepackage{dcolumn}
\usepackage{bm}
\usepackage{graphicx,epsfig}

\input{tcilatex}

\begin{document}

\author{Tao Xu$^{1,2}$\thanks{%
{} E-mail address: xutao@hust.edu.cn, }, and Richard Fitzpatrick$^{1}$,
Francois L. Waelbroeck$^{1}$}
\title{Vacuum Solution for Solov'ev's Equilibrium Configuration in Tokamaks}
\affiliation{$^{1}$Institute for Fusion Studies, University of Texas at Austin, Austin,
TX 78712, USA. \\
$^{2}$Institute of Fusion and Plasma Research, College of Electrical and
Electronic Engineering, Huazhong University of Science and Technology, Wuhan
430074, PRC. }

\begin{abstract}
In this work, we construct a simple tokamak plasma equilibrium generated by
currents flowing within the plasma and currents flowing in distant external
coils. The plasma current density takes the form $j_{\varphi
}(r,z)=-ar-bR^{2}/r$ inside the plasma, and is zero in the surrounding
vacuum. We use Green's function method to compute the plasma current
contribution, together with a homogeneous solution to the Grad-Shafranov
equation, to construct the full solution. Matching with the constant
boundary condition on the last closed flux surface is performed to determine
the homogeneous solution. The total solution is then extended into the
vacuum region to get a realistic vacuum solution. We find that the actual
solution is different from the Solov'ev solution, especially the X-point
structure. The X-point obtained at the last closed flux surface is not like
the letter "X", and the expanded angle in the vacuum is larger than
corresponding angle in the plasma at the null point. The results are
important for understanding the X-point and separatrix structure. At the end
of the paper, we have extended the classic Solovev's configuration to an
ITER-like configuration, and obtained the vacuum solution.

PACS number(s): 52.55.-s, 47.65.-d.

\end{abstract}

\maketitle


\section{\protect\bigskip Introduction}

Equilibrium computation is crucial for the design and operation of magnetic
fusion devices. The equilibrium magnetic configuration in magnetic
confinement devices is determined by the Grad-Shafranov (GS) equation [$1$-$%
3 $]. The Grad-Shafanov equation is a nonlinear elliptic partial
differential equation, which is usually solved by numerical computation.

In 1968, Solov'ev \cite{Solov} proposed simple linear stream functions, and
got analytic solution for the Grad-Shafranov equation. Solov'ev's
equilibrium configurations are useful for the benchmarking
magnetohydrodynamics equilibrium codes \cite{Johnson,Lao}, as well as
stability analysis \cite{Green} of toroidal axisymmetric tokamaks. They have
been used to construct up--down symmetric tokamak equilibria \cite{Weening},
as well as a divertor configuration. Recently, many extended analytic works [%
$9$-$11$] including a single-null solution to the GS equation are presented,
which can be used in different situations for fusion devices.

Solov'ev's theory assumes there is distributed current filling all space,
including the vacuum region. But, in fact, the plasma current density is
almost zero in the vacuum region. Even in the region between last closed
flux surface (LCFS) and limiter or divertor, plasma current behavior is very
different to behavior inside the LCFS \cite{St}. So, Solov'ev's vacuum
solution is inapplicable. The full vacuum solution of Grad-Shafrnov equation
has two parts. The first part is the contribution from the plasma currents.
The second part is the contribution from the external current. In order to
construct the full vacuum solution, we use the Green's function method to
compute the plasma current contribution, together with the homogeneous
solution to the Grad-Shafranov equation. We have matched the plasma-vacuum
boundary for the last closed flux surface. Hence, we get a realistic vacuum
solution for Solov'ev's equilibrium configuration in this paper.

This paper is organized as follows. The problem of Solov'ev's solution is
discussed in Section $2$. Our method to solve the vacuum field in fusion
devices is presented in Section $3$. The simulation results obtained by our
method are presented in this section. Vacuum solution of ITER-like
configurations are discussed in Section $4$. The main conclusions are
summarized in Section $5$.

\section{Solov'ev's solution and its problem}

\ We use cylindrical coordinates $r,\varphi ,z$ to describe toroidally
axisymmetric plasma configurations. It is well known that the poloidal flux
function $\psi (r,z)$ satisfies the Grad-Shafranov equation%
\begin{equation}
\frac{\partial ^{2}\psi }{\partial z^{2}}+r\frac{\partial }{\partial r}\frac{%
1}{r}\frac{\partial \psi }{\partial r}=-rj_{\varphi }\text{,}  \label{1}
\end{equation}%
where $j_{\varphi }$ is the longitudinal current density along the cross
section of the plasma. $j_{\varphi }$ is determined by relation%
\begin{equation}
\qquad j_{\varphi }=rp^{\prime }+\frac{I_{A}I_{A}^{\prime }}{r}\text{,}
\label{2}
\end{equation}%
where $p$ is the plasma pressure, and $I_{A}$ is the net poloidal current
function. Both $p$ and $I_{A}$ are free functions of $\psi $. The prime
denotes differentiation with respect to $\psi $.

The magnetic field $\mathbf{B}$ can be expressed as 
\begin{equation}
\mathbf{B}=\frac{I_{A}}{r}\mathbf{e}_{\varphi }+\frac{1}{r}\nabla \psi
\times \mathbf{e}_{\varphi },  \label{3}
\end{equation}

The poloidal flux function $\psi (r,z)$, the pressure functions $-p^{\prime
} $ and $-I_{A}I_{A}^{\prime }$ can be expanded in a power series in $%
r^{2}-R^{2}$ and $z$. If we keep first $3$rd degree polynomials of $%
r^{2}-R^{2}$ and $z$ with up-down symmetry, i.e., $r^{2}-R^{2}$, $%
(r^{2}-R^{2})^{2}$, $z^{2}$, $(r^{2}-R^{2})^{3}$, $(r^{2}-R^{2})z^{2}$, we
substitute them into Equation (\ref{1}), and obtain the Solov'ev
configuration equation 
\begin{equation}
\psi =\frac{AR^{2}}{2}(1+C\frac{r^{2}-R^{2}}{R^{2}})z^{2}+\frac{a+b-A}{8}%
(r^{2}-R^{2})^{2}-\frac{b-(1-C)A}{24R^{2}}(r^{2}-R^{2})^{3}\text{.}
\label{10}
\end{equation}%
The cross sections of the magnetic surfaces $\psi =$const near the magnetic
axis ($r=R$, $z=0$) are ellipses. The ratio of their semi-radii $%
l_{z}/l_{r}=\epsilon $ determines the constant $A=(a+b)/(1+\epsilon ^{2})$.
In special case that $-p^{\prime }=a$ and $-I_{A}I_{A}^{\prime }=bR^{2}$,
Solov'ev got several interesting configurations in his paper \cite{Solov}.

A distributed current filling all space, including the vacuum region, is
assumed in Solov'ev's theory. But, in fact, the plasma current density is
almost zero in vacuum region of fusion devices. So, Solov'ev's solution is
inapplicable in vacuum region. In next section, we will give a realistic
computation for the vacuum solution of Solov'ev's equilibrium configuration
in fusion devices.

\section{Our method}

The full vacuum solution of Grad-Shafrnov equation has two parts. The first
part is the contribution from the plasma currents. We will use the Green's
function method to compute the plasma current contribution. The second part
is the contribution from the external coils far away the plasma region, and
can be expanded by Taylor series. It is a homogeneous solution to the
Grad-Shafranov equation.

In order to calculate $\psi $, we let 
\begin{equation}
\zeta =\frac{r^{2}-R^{2}}{2R},\qquad \tan \theta =\frac{z}{\zeta }\text{.}
\label{11}
\end{equation}%
Then equation ($5$) can be written as%
\begin{equation}
\psi =(ACR\tan ^{2}\theta -\frac{b-(1-C)AR}{3})\zeta ^{3}+(\frac{1}{2}%
AR^{2}\tan ^{2}\theta +\frac{a+b-A}{2}R^{2})\zeta ^{2}\text{.}  \label{12}
\end{equation}

Now, 
\begin{equation}
\frac{\partial (\psi ,\theta )}{\partial (\zeta ,z)}\equiv \left\vert 
\begin{array}{cc}
\frac{\partial \psi }{\partial \zeta }|_{z} & \frac{\partial \psi }{\partial
z}|_{\zeta } \\ 
\frac{\partial \theta }{\partial \zeta }|_{z} & \frac{\partial \theta }{%
\partial z}|_{\zeta }%
\end{array}%
\right\vert \text{.}  \label{17}
\end{equation}

We can get%
\begin{eqnarray}
\frac{\partial (\psi ,\theta )}{\partial (\zeta ,z)}
&=&[ACRz^{2}+(a+b-A)R^{2}\zeta -[b-(1-C)A]R\zeta ^{2}]\frac{\zeta }{\zeta
^{2}+z^{2}}  \notag \\
&&+(2ACR\zeta z+AR^{2}z)\frac{z}{\zeta ^{2}+z^{2}}\text{.}  \label{19}
\end{eqnarray}%
It follows that%
\begin{equation}
\frac{\partial (\psi ,\theta )}{\partial (\zeta ,z)}=AR^{2},\qquad \text{if
\ \ \ }\theta =\frac{\pi }{2}(2n-1)\text{, }  \label{20}
\end{equation}%
and%
\begin{equation}
\frac{\partial (\psi ,\theta )}{\partial (\zeta ,z)}%
=(a+b-A)R^{2}-[b-(1-C)A]R\zeta \text{, \ \ if \ \ \ }\theta =\pi n\text{,}
\label{23}
\end{equation}%
where $n$ is an integer. The magnetic surface function $\psi _{G}$
calculated by Green's function method is 
\begin{equation}
\psi _{G}(r,z)=\int_{JR}G(r,z;r^{\prime },z^{\prime })j_{\varphi }(r^{\prime
},z^{\prime })dr^{\prime }dz^{\prime }\text{,}  \label{34}
\end{equation}%
where the integral is over the plasma current area. The Green's function is
given by \cite{Jacs}%
\begin{equation}
G(r,z;r^{\prime },z^{\prime })=\frac{1}{2\pi }\frac{\sqrt{rr^{\prime }}}{k}%
[(2-k^{2})K(k)-2E(k)]\text{,}  \label{Grenn}
\end{equation}%
where\ \ \ \ \ \ \ \ \ \ \ \ \ \ \ \ \ \ \ \ \ \ \ \ \ \ \ \ \ \ \ \ \ \ \ \
\ \ \ \ \ \ \ \ \ \ \ \ \ \ \ \ \ \ \ \ \ \ \ \ \ \ \ 
\begin{equation}
k^{2}=\frac{4rr^{\prime }}{\sqrt{(r+r^{\prime })^{2}+(z-z^{\prime })^{2}}}%
\text{.}  \label{k}
\end{equation}

The external coils generate homogeneous solution, which satisfies 
\begin{equation}
\frac{\partial ^{2}\psi }{\partial z^{2}}+r\frac{\partial }{\partial r}\frac{%
1}{r}\frac{\partial \psi }{\partial r}=0\text{.}  \label{35}
\end{equation}

The homogeneous solution can be written as%
\begin{equation}
\psi _{h}=c_{1}\psi _{h0}+c_{2}\psi _{h2}+c_{3}\psi _{h4}+c_{4}\psi _{h6}+...
\label{36hc}
\end{equation}%
where \cite{Za,Reu,Jardin}%
\begin{equation}
\psi _{h0}=R^{2}  \label{36h1}
\end{equation}

\begin{equation}
\psi _{h2}=\frac{1}{2}(r^{2}-R^{2})  \label{36h2}
\end{equation}

\begin{equation}
\psi _{h4}=\frac{1}{8R^{2}}[(r^{2}-R^{2})^{2}-4r^{2}z^{2}]  \label{36h3}
\end{equation}

\begin{equation}
\psi _{h6}=\frac{1}{24R^{4}}%
[(r^{2}-R^{2})^{3}-12r^{2}z^{2}(r^{2}-R^{2})+8r^{2}z^{4}]\text{.}
\label{36h4}
\end{equation}%
\begin{equation}
\psi _{h8}=\frac{1}{320R^{6}}%
[5(r^{2}-R^{2})^{4}-120r^{2}z^{2}(r^{2}-R^{2})^{2}+80r^{2}z^{4}(3r^{2}-2R^{2})-64r^{2}z^{6}]%
\text{.}  \label{36h5}
\end{equation}%
The above terms are corresponding to the even nullapole, dipole, quadrupole
and hexapole because Solov'ev's solution satisfies up-down symmetry.

The inhomogeneous poloidal flux function $\psi $ is obtained from Equation (%
\ref{34}), where 
\begin{equation}
drdz=\frac{R}{\sqrt{R^{2}+2R\zeta }}\frac{d\psi d\theta }{\frac{\partial
(\psi ,\theta )}{\partial (\zeta ,z)}}\text{.}  \label{cor}
\end{equation}

The complete flux function is thus,%
\begin{eqnarray}
\psi (r,z) &=&\int_{jR}G(r,z;r^{\prime },z^{\prime })j_{\varphi }(r^{\prime
},z^{\prime })dr^{\prime }dz^{\prime }+\psi _{h}  \notag \\
&=&\int_{jR}G(r,z;\psi ,\theta )j_{\varphi }(\psi ,\theta )\frac{R}{\sqrt{%
R^{2}+2R\zeta }}\frac{\partial (\zeta ,z)}{\partial (\psi ,\theta )}d\psi
d\theta +\psi _{h}  \label{41} \\
&\approx &{\displaystyle \sum} {\displaystyle \sum} G(r,z;\psi ,\theta
)j_{\varphi }(\psi ,\theta )\frac{R}{\sqrt{R^{2}+2R\zeta }}\frac{\partial
(\zeta ,z)}{\partial (\psi ,\theta )}\delta \psi \delta \theta +\psi _{h}%
\text{.}  \notag
\end{eqnarray}

The Green function solution $\int_{jR}G(r,z;r^{\prime },z^{\prime
})j_{\varphi }(r^{\prime },z^{\prime })dr^{\prime }dz^{\prime }$ is the full
solution for plasma current in the whole space. The homogeneous solution $%
\psi _{h}$ is the full solution for external coils, which is used to satisfy
the plasma-vacuum boundary condition. They are calculated by Biot-Savart
law. The poloidal flux function satisfies that $\psi =-rA_{\varphi }$ where $%
A_{\varphi }$ is the longitudinal component of vector potential. The
corresponding magneic induction $B_{r}=-\frac{1}{r}\frac{\partial \psi }{%
\partial z}$, $B_{z}=\frac{1}{r}\frac{\partial \psi }{\partial r}$.
Therefore, if the solution of Equation (\ref{41}), $\psi (r,z)$ or $\mathbf{B%
}$, satisfies the boundary condition on the LCFS, due to the solution
uniqueness \cite{Fit} one concludes that Equation (\ref{41}) is the solution
of the system. However, since Equation (\ref{41}) is just the Biot-Savart
law, it can be extended to the vacuum. There are no discontinuity issues
across the plasma-vacuum interface. The $\psi (r,z)$ or $\mathbf{B}$ is
automatically continuous. This method is described in the references [$5$]
to compute the vacuum solution from the fixed boundary solution.

If the plasma current is given, the value of $\int_{jR}G(r_{b},z_{b};r^{%
\prime },z^{\prime })j_{\varphi }(r^{\prime },z^{\prime })dr^{\prime
}dz^{\prime }$ is given. $\psi _{b}(r_{b},z_{b})$ is the value of Solov'ev
solution on the LCFS, which is fixed too. We denote 
\begin{equation}
\delta \psi _{b}(r_{b},z_{b})=\psi
_{b}(r_{b},z_{b})-\int_{jR}G(r_{b},z_{b};r^{\prime },z^{\prime })j_{\varphi
}(r^{\prime },z^{\prime })dr^{\prime }dz^{\prime }\text{.}  \label{bound2}
\end{equation}%
If $\psi _{h}(r_{b},z_{b})$, i.e.,${\displaystyle\sum }_{i}c_{i}\psi _{hi}$,
is exact the value $\delta \psi _{b}(r_{b},z_{b})$, due to the solution
uniqueness $\psi _{2}(r,z)$ is the same solution as the Solov'ev solution in
the plasma region. The extension of $\psi _{2}(r,z)$ to the vacuum is the
vacuum solution. But, we can not get exact $\psi _{h}(r_{b},z_{b})$ by
analytic method, and we just use the combination of multipoles to
approximate $\psi _{h}(r_{b},z_{b})$. The infinite mutipoles can infinitely
approximate the $\delta \psi _{b}(r_{b},z_{b})$. Here we use nullapole,
dipole, quadrupole, hexapole, and octopole to fit the LCFS. Then there are
only five parameters are undetermined, this leads us to match the boundary
for plasma region solution in five typical points. In order to determine
unknown $c_{n}$, we have selected observation points a small distance $%
\epsilon $\ outside the plasma-vacuum boundary, and take the limit as $%
\epsilon \rightarrow 0$. A good choice is to match boundary condition near
five characteristic points: the inner equatorial point with $r^{\prime
}=r_{\min }^{\prime }$, $z^{\prime }=0$; the outer equatorial point with $%
r^{\prime }=r_{\max }^{\prime }$, $z^{\prime }=0$; the highest vertically
point of the LCFS; the outer point in the last closed flux surface with $%
r^{\prime }=R$ \cite{Fre1} and another point. These five observation points
should have the same poloidal flux. Hence, we get the value of $c_{n}$ and
the full solution of $\psi (r,z)$ (\ref{41}) by this method.

Let us see the first case: $a=1.2$, $b=-5a/6$, $C=11$, $A=a/12$, $R=10$. The
magnetic surface function in the plasma region is 
\begin{equation}
\psi =5(1+11\frac{r^{2}-100}{100})z^{2}+\frac{1}{80}(r^{2}-100)^{2}\text{.}
\label{46}
\end{equation}

The corresponding Solov'ev configuration is shown in figure $1$. The above
configuration is composed by $z^{2}$, $(r^{2}-R^{2})z^{2}$, $%
(r^{2}-R^{2})^{2}$. So we just have used the nullapole, dipole, quadrupole
to fit the above configuration. The three characteristic points: the inner
equatorial point with $r^{\prime }=r_{\min }^{\prime }$, $z^{\prime }=0$;
the outer equatorial point with $r^{\prime }=r_{\max }^{\prime }$, $%
z^{\prime }=0$; the highest vertically point of the LCFS. The poloidal flux
function distribution calculated by our method is shown in figure $2$. The
shape of the X-point with $\psi =1.033057$ calculated by our method is
compared with X-point configuration of Solov'ev's solution in figure $3$.
The dotted lines are the X-points configurations obtained by our method, and
the solid lines are the X-points configurations of Solov'ev's solution. The
shape of the X-points configurations in our method are different from
Solov'ev's result. The vertical angles of X-points of Solov'ev's solution
are always equal and the X-point obtained in Solov'ev's solution is exactly
the letter "X". It is shown by solid line in figure $3$. But the vertical
angles obtained by our method are not equal, and the X-point obtained by our
method is not like the letter "X". The expanded angle $\Theta _{V}$ in the
vacuum is larger than corresponding angle $\Theta _{p}$ in the plasma for
our solution, i.e. $\Theta _{V}>\Theta _{P}$. It is shown by dotted line in
figure $3$. The magnitude of the $\Theta _{V}-\Theta _{P}$ is much larger
than our calculation deviation.

\section{Vacuum solution of ITER-like configuration}

The Solov'ev configuration has up-down symmetry, but the magnetic
configuration in ITER does not have up-down symmetry. We will extend the
classic Solov'ev configuration to an ITER-like configuration with up-down
asymmetry. In order to get an ITER-like configuration, we have extended
Solov'ev solution (\ref{10}) to this form 
\begin{eqnarray}
\psi &=&a_{1}(r^{2}-R^{2})+b_{1}z+c_{1}(r^{2}-R^{2})z+\frac{AR^{2}}{2}(1+C%
\frac{r^{2}-R^{2}}{R^{2}})z^{2}+  \notag \\
&&\frac{a+b-A}{8}(r^{2}-R^{2})^{2}-\frac{b-(1-C)A}{24R^{2}}(r^{2}-R^{2})^{3}%
\text{.}  \label{ITER1}
\end{eqnarray}%
Here, $a_{1}(r^{2}-R^{2})$ is a translation term, $b_{1}z$ and $%
c_{1}(r^{2}-R^{2})z$ are up-down asymmetry terms. It is obvious that the
above equation satisfies GS equation with $-p^{\prime }=a$ and $%
-I_{A}I_{A}^{\prime }=bR^{2}$.

Let us consider the case that $a_{1}=0.041$, $b_{1}=0.091,$ $c_{1}=-0.02$, $%
a=1$, $b=-9a/11$, $C=10$, $A=a/11$, $R=10$. The poloidal magnetic surface
function in plasma region is 
\begin{equation}
\psi =0.041(r^{2}-R^{2})+0.091z-0.02(r^{2}-R^{2})z+\frac{50}{11}(1+10\frac{%
r^{2}-100}{100})z^{2}+\frac{1}{88}(r^{2}-100)^{2}\text{.}  \label{Iter}
\end{equation}

The analytic solution (\ref{Iter}) of the magnetic surface function is shown
in figure $4$. It is an ITER-like configuration with a divertor
configuration, whose vacuum solution can be written as%
\begin{equation}
\psi =\psi _{G}+\psi _{h}\text{.}  \label{ITER2}
\end{equation}%
Here $\psi _{G}$ is the vacuum solution by plasma current density, $\psi
_{h} $ is the homogeneous solution. The homogeneous solution can be written
as%
\begin{equation}
\psi _{h}=C_{1}\psi _{h0}+C_{2}\psi _{h2}+C_{3}\psi _{h4}+C_{4}\psi
_{h6}+C_{5}\psi _{h8}+C_{6}\psi _{h3}+C_{7}\psi _{h5}+...  \label{ITER4}
\end{equation}%
where \cite{Jardin}%
\begin{eqnarray}
\psi _{h3} &=&\frac{1}{R}r^{2}z,  \notag \\
\psi _{h5} &=&\frac{1}{6R^{3}}r^{2}z[3(r^{2}-R^{2})-4z^{2}]\text{.}
\label{ITER5} \\
\psi _{h7} &=&\frac{1}{60R^{5}}%
[15r^{2}z(r^{2}-R^{2})^{2}+20r^{2}z^{3}(2R^{2}-3r^{2})+24r^{2}z^{5}]\text{.}
\notag
\end{eqnarray}

We have not consider the factor $\psi _{h1}$ because $\psi _{h1}=0$. $\psi
_{h3},\psi _{h5},\psi _{h7}$ are first few odd multipoles, and $\psi _{h3}$
have the terms ($r^{2}-R^{2})z$ and $z$. So we use $\psi _{h0}$, $\psi _{h2}$%
, $\psi _{h4}$, $\psi _{h6}$, $\psi _{h8}$, $\psi _{h3}$, $\psi _{h5}$,$\psi
_{h7}$ to fit ITER-like Solov'ev configuration (\ref{ITER1}), and need $8$
coefficients. In order to obtain the $8$ coefficients, we have selected
seven points very close to the LCFS including the $4$ extreme values: ($%
r_{\max }$,$0$), ($r_{\min }$,$0$), ($r$,$z_{\max }$), and lowest X-point $%
(r_{s},z_{s})$. So, we obtain the vacuum solution of this ITER-like
configuration, which is shown in figure $5$. The first X-point is plotted by
black line with $\psi =0.5471$. If an X-point is at the plasma-vacuum
boundary, the expanded vertical angles $\Theta _{V}$ in vacuum region are
larger than the corresponding ones $\Theta _{p}$ in plasma region in our
cases. The vertical angle difference is shown clearly in figure $6$. The
second X-point has appeared at $\psi =0.5472$, which is shown by red line in
figure $5$.

The poloidal magnetic surface function satisfies $\frac{\partial \psi }{%
\partial r}=\frac{\partial \psi }{\partial z}=0$ at X-points ($r_{s},z_{s}$%
). The classical Solov'en solution (\ref{10}) has up-down symmetry, so we
have 
\begin{equation}
\psi \approx \psi (r_{s},z_{s})+\frac{1}{2}\frac{\partial ^{2}\psi }{%
\partial r^{2}}|_{(r_{s},z_{s})}(r-r_{s})^{2}+\frac{1}{2}\frac{\partial
^{2}\psi }{\partial z^{2}}|_{(r_{s},z_{s})}(z-z_{s})^{2}\text{.}  \label{XP3}
\end{equation}%
Let us consider the points: $(r_{p},z_{p})$ inside the LCFS, and $%
(r_{V},z_{V})$ outside the LCFS. These points are close to the X-point ($%
r_{s},z_{s}$) at the plasma-vacuum boundary. Then $\frac{\partial ^{2}\psi }{%
\partial z^{2}}\mid _{(r_{p},z_{p})}+\frac{\partial ^{2}\psi }{\partial r^{2}%
}\mid _{(r_{p},z_{p})}\approx ar_{p}^{2}+bR^{2}$, $\frac{\partial ^{2}\psi }{%
\partial z^{2}}\mid _{(r_{V},z_{V})}+\frac{\partial ^{2}\psi }{\partial r^{2}%
}\mid _{(r_{V},z_{V})}\approx 0$. We can also take Taylor's expansion to $%
\frac{\partial ^{2}\psi }{\partial z^{2}}\mid _{(r_{V},z_{V})}$ and $\frac{%
\partial ^{2}\psi }{\partial r^{2}}\mid _{(r_{V},z_{V})}$ close to the
X-point. It is obvious that $\psi =\psi (r_{s},z_{s})$ for the LCFS. Then $%
\frac{z-z_{s}}{r-r_{s}}\approx \pm \sqrt{-\frac{\partial ^{2}\psi }{\partial
r^{2}}|_{(r_{s},z_{s})}/\frac{\partial ^{2}\psi }{\partial z^{2}}%
|_{(r_{s},z_{s})}}$ in the LCFS near the X point. So the expanded angle $%
\Theta _{V}$ in vacuum region is almost $90$ degree. Plasma current density $%
j_{\varphi }(r_{s},z_{s})$, i.e. $ar_{p}^{2}+bR^{2}$ does not equal zero
usually in the plasma region. So it means that expanded angles $\Theta
_{V}\neq $ $\Theta _{p}$.

Let us consider the extended Solov'ev solution (\ref{ITER1}). The poloidal
magnetic surface function near the X-point is 
\begin{equation}
\psi \approx \psi (r_{s},z_{s})+\frac{1}{2}\frac{\partial ^{2}\psi }{%
\partial r^{2}}|_{(r_{s},z_{s})}(r-r_{s})^{2}+\frac{1}{2}\frac{\partial
^{2}\psi }{\partial z^{2}}|_{(r_{s},z_{s})}(z-z_{s})^{2}+\frac{\partial
^{2}\psi }{\partial r\partial z}|_{(r_{s},z_{s})}(r-r_{s})(z-z_{s})\text{.}
\label{XP5}
\end{equation}%
We define $\tau _{1}=\frac{\partial ^{2}\psi }{\partial r^{2}}%
|_{(r_{s},z_{s})}$, $\tau _{2}=\frac{\partial ^{2}\psi }{\partial z^{2}}%
|_{(r_{s},z_{s})}$, $\tau _{12}=\frac{\partial ^{2}\psi }{\partial r\partial
z}|_{(r_{s},z_{s})}$. The expanded angle is about arctan$\frac{\sqrt{\tau
_{12}^{2}-4\tau _{1}\tau _{2}}}{\tau _{1}+\tau _{2}}$. It is obvious that
expanded angles $\Theta _{V}\neq $ $\Theta _{p}$.

\section{Summary}

In this paper, the vacuum solution for Solov'ev's equilibrium configuration
in fusion devices is presented and the Solov'ev formulation is inapplicable
in vacuum region. We use the Green's function method to compute the plasma
current contribution, together with the homogeneous solution to the
Grad-Shafranov equation, to construct the vacuum solution. Moreover, we have
extended the classic up-down Solov'ev's configuration to an ITER-like
configuration with non up-down symmetric equilibria.

We find that the actual vacuum solution is vastly different from the
Solov'ev solution in the vacuum region, especially the X-point structure.
The X-point obtained at the last closed flux surface is not like the letter
"X", and the expanded angle in the vacuum is larger than corresponding angle
in the plasma. These results are obtained first time in the world. The
unsymmetrical current density gradient causes inequality of vertical angles
near the X-point at plasma-vacuum boundary. The value difference of vertical
angles, i.e. $\Theta _{V}-\Theta _{P}$ is generated due to the difference of
current distribution between the plasma region and vacuum region. It is
expected to be tested by the divertor tokamaks in future.

The equilibrium magnetic field topology in a divertor tokamak configuration
consists of closed nested torus inside LCFS and open field line outside.
Many particles move along the magnetic field. So the open field line outside
has determined the deposition of particles and energy. The distance from the
X-point to the divertor is fixed. The expanded angles in the vacuum is the
same with the ones in the plasma according to the classical equilibrium
calculation. If the expanded angles facing the divertor can be larger than
the ones in the plasma for the LCFS, the particles deposition area and
particles distribution volume are more larger. So the particles distribution
and energy deposition will be loosen than results before. The rapid
deposition of energy onto plasma facing components by edge localized modes
ELMs is a potentially serious impediment to the production of fusion energy
in large tokamaks, such as ITER \cite{Aymar} or DEMO \cite{DEMO}. Based on
our results, we make a conclusion that the expanded angles facing Scrape-Off
Layer can be larger than the corresponding ones in plasma region. The value
difference of vertical angles can be varied by divertor control, and the
particles density of deposition can be looser or tighter by different
control methods.

In addition, extra X-points occur sometimes far away the magnetic aixs by
this method. These results are important for understanding the X-point and
separatrix structure.

The method in this paper can be used to solve the vacuum solution for the GS
equation in an arbitrarily shaped, finite-pressure toroidal plasma of fusion
devices. The current density distribution can be extended to construct
symmetric and asymmetric divertor configurations. The homogeneous solution
defined in GS equation (\ref{35}) contains symmetric and asymmetric terms (%
\ref{ITER5}). This can be used to construct up--down symmetric and
asymmetric equilibria with specified null-point location.

\bigskip

\textbf{Acknowledge}

This work was supported by the U.S. Department of Energy under Contract No.
DE-FG02-04ER-54742. The author Tao Xu thanks the fundation supported by
China Scholarship Council and ITER special fundation of China under Contract
No. 2017YFE0301202.

\end{document}